\documentclass[review]{elsarticle}

\usepackage{booktabs}
\usepackage{multirow}
\usepackage{graphicx}
\usepackage{subfigure}
\usepackage{comment}
\usepackage{color}
\usepackage{algorithm}
\usepackage{algorithmic}
\usepackage{setspace}

\journal{Journal of \LaTeX\ Templates}

\biboptions{authoryear}

\begin{document}

\begin{frontmatter}

\title{EOCSA: Predicting Prognosis of Epithelial Ovarian Cancer with Whole Slide Histopathological Images}

\author[mymainaddress]{Tianling Liu}
\ead{dling@tju.edu.cn}
\author[mymainaddress]{Ran Su\corref{mycorrespondingauthor}}
\ead{ran.su@tju.edu.cn}
\author[changmingaddress]{Changming Sun}
\ead{changming.sun@csiro.au}

\author[xiutingmymainaddress]{Xiuting Li}
\ead{li\_xiuting@sbic.a-star.edu.sg}

\author[mysecondaryaddress]{Leyi Wei\corref{mycorrespondingauthor}}
\ead{weileyi@sdu.edu.cn}
\cortext[mycorrespondingauthor]{Corresponding author}
%\ead{ran.su@tju.edu.cn and weileyi@tju.edu.cn}

\address[mymainaddress]{School of Computer
	Software, College of Intelligence and Computing, Tianjin University, China.}
\address[mysecondaryaddress]{School of Software, Shandong University, China.}
\address[changmingaddress]{CSIRO Data61, Epping, NSW 1710, Australia.}% <-this % stops a space
\address[xiutingmymainaddress]{Singapore Bio-imaging Sciences Consortium (SBIC), Agency for Science, Technology and Research (A*STAR), Singapore.}% <-this % stops a space

\begin{abstract}
Ovarian cancer is one of the most serious cancers that threaten women around the world. Epithelial ovarian cancer (EOC), as the most commonly seen subtype of ovarian cancer, has rather high mortality rate and poor prognosis among various gynecological cancers. Survival analysis outcome is able to provide treatment advices to doctors. In recent years, with the development of medical imaging technology, survival prediction approaches based on pathological images have been proposed. In this study, we designed a deep framework named EOCSA which analyzes the prognosis of EOC patients based on pathological whole slide images (WSIs). Specifically, we first randomly extracted patches from WSIs and grouped them into multiple clusters. Next, we developed a survival prediction model, named DeepConvAttentionSurv (DCAS), which was able to extract patch-level features, removed less discriminative clusters and predicted the EOC survival precisely. Particularly, channel attention, spatial attention, and neuron attention mechanisms were used to improve the performance of feature extraction. Then patient-level features were generated from our weight calculation method and the survival time was finally estimated using LASSO-Cox model. The proposed EOCSA is efficient and effective in predicting prognosis of EOC and the DCAS ensures more informative and discriminative features can be extracted. As far as we know, our work is the first to analyze the survival of EOC based on WSIs and deep neural network technologies. The experimental results demonstrate that our proposed framework has achieved state-of-the-art performance of 0.980 C-index. The implementation of the approach can be found at https://github.com/RanSuLab/EOCprognosis.
\end{abstract}


\end{frontmatter}

\section{Introduction}
Ovarian cancer is a disease that seriously damages the well-being of women in the world. It ranks the seventh among the most common cancers in women and the eighth among all female deadly cancers~\citep{lheureux2019epithelial}. Epithelial ovarian cancer (EOC), as the most commonly seen subtype of ovarian cancer, accounts for more than 90\% among all ovarian cancer patients~\citep{reid2017epidemiology,torre2018ovarian}. Although the 5-year relative survival rate at the early stage of EOC is as high as 93\%~\citep{torre2018ovarian}, the symptoms at the early stage of EOC are not obvious, which make EOC difficult to be detected at the early stage~\citep{lheureux2019epithelial}. In contrast, more than 75\% of EOCs are detected at an advanced stage and the 5-year relative survival rate of EOC at an advanced stage is only about 29\%~\citep{lheureux2019epithelial,baal2018development}.
%%remove zhu2016lung and haarburger2019imagebased after of diseases
Survival analysis is a branch of statistics which concentrates on predicting the time duration from the beginning of the follow-up study to the interest events occur, for example, disease recurrence or death~\citep{ohnomachado2001modeling}. An important application of survival analysis in the medical field is to observe the prognosis of diseases~\citep{huang2018deep,kather2019predicting}. Survival analysis can make a rough prediction of patients' survival status over time, and can also roughly assess the development trend of the disease condition. There are many factors in EOC patients' data, such as age, gender, pathogenesis and tumor stage~\citep{holschneider2000ovarian,tingulstad2003survival}. Survival analysis can determine which factor or set of factors has a greater impact on patient's survival, so that doctors can evaluate the effectiveness of different treatment plans to tailor for each patient. Doctors can also change the treatment plans or drugs according to the results of survival analysis during the treatment of patients to ensure that patients have a high survival rate.
%%remove yang2015evaluation after proposed
With the development of medical imaging technology, various types of cancer images are available such as MRIs, CT images and pathological images. These high-quality images contain rich information related to the characteristics of cancers, thus a large number of image-based survival analysis methods have been proposed~\citep{wang2014novel,Yu2016Predicting,Lu2019A}. Pathological images, especially the whole slide images (WSIs) have been adopted to perform survival analysis and have shown impressive performance~\citep{wang2014novel,Yu2016Predicting,zhu2016deep}. Most of these methods firstly extracted hand-crafted features and then used machine learning algorithms to conduct survival prediction~\citep{wang2014novel,Yu2016Predicting}. However, the hand-crafted features extracted from the WSIs requires predefinition of features, thus the performance largely depends on prior knowledge which may bring bias into the results.

The development of deep learning brings about the automated acquisition of features and has increasingly demonstrated the ability to extract high-dimensional features~\citep{DUNet2019,Method2019}. In recent years, there are many survival analysis studies based on deep learning. Zhu et al. proposed the DeepConvSurv model which applied deep convolutional neural network to image-based survival analysis tasks~\citep{zhu2016deep}. For further solving WSI-based survival analysis tasks, Zhu et al. proposed the WSISA framework that estimated the survival using discriminative patches extracted from WSIs~\citep{zhu2017wsisa}. Mobadersany et~al. integrated WSIs and genetic data to make cancer outcomes prediction using convolutional neural network (CNN)~\citep{mobadersany2018predicting}. Using WSIs and deep learning to analyze EOC survival still requires exploration.

In this study, inspired by the WSISA, we proposed a EOCSA framework for the analysis of EOC survival based on WSI data. Firstly, we randomly picked a certain number of patches from each WSI and grouped them into different clusters. Then we trained a deep survival prediction model named DeepConvAttentionSurv (DCAS) for each cluster. Next, we selected the clusters which contained patches with discriminative information and extracted the patch-level features based on the trained DCAS models. Finally, we generated weighted patient-level features based on patch-level features to make final survival analysis. Compared to the WSISA, the EOCSA is able to extract more effective features and make more accurate prediction. We have three contributions in this study. Firstly, as far as we know, we are the first to propose the use of deep learning to process WSIs for survival analysis of EOC; Secondly, we developed an efficient deep survival analysis model named DCAS. In order to extract more effective features, we added spatial, channel and neuron attention modules into the architecture; Thirdly, we proposed a new weight calculation method to obtain more discriminative patient-level features compared with the existing methods. The last two contributions are algorithmic innovations. And the first contribution is application innovation. We apply two algorithmic innovation to the survival analysis of EOC.  Experiments demonstrate the efficiency and effectiveness of the proposed EOCSA in predicting EOC patients' survival. The implementation of the approach can be found at https://github.com/RanSuLab/EOCprognosis.

\begin{figure*}
	\centering
	\includegraphics[width=4.7in]{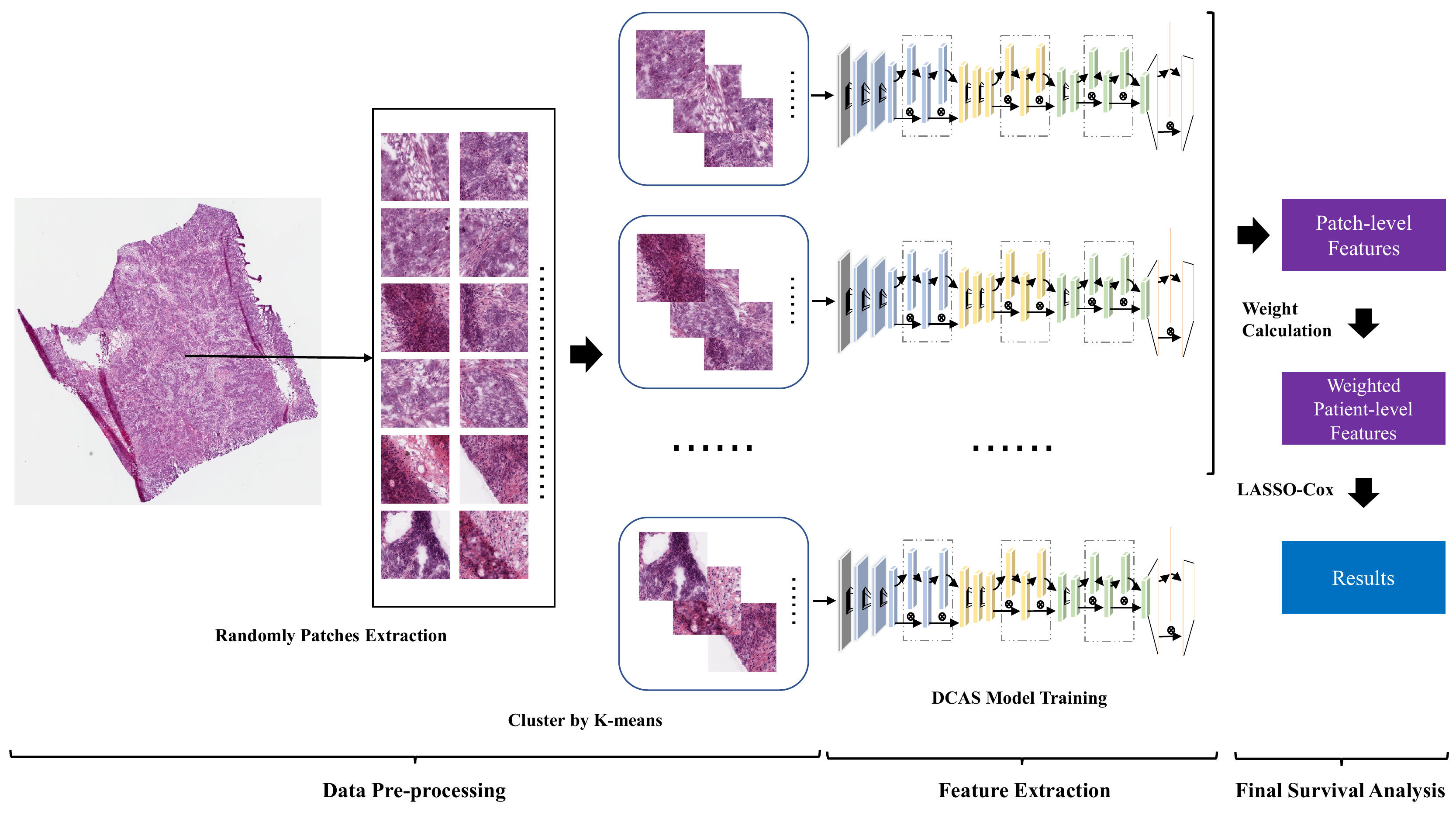}
	\caption{The overview of our EOCSA framework. There are three main steps in EOCSA: 1. Data pre-processing, model training; 2. Feature extraction; 3. Weighted patient-level feature generation and survival prediction.}
	\label{EOCSA}
\end{figure*}

\section{Background}
Survival analysis is frequently used to observe the prognosis of medical diseases. The Kaplan-Meier survival estimate, a non-parameter method, predicts survival rate by observed survival time~\citep{dabrowska1988kaplanmeier}. It is the easiest way to calculate survival rate over time~\citep{goel2010understanding}. Mevlut et al. combined Kaplan-Meier analysis with decision trees to predict survival of breast cancer~\citep{turemevlut2009using}. But the drawback of Kaplan-Meier survival estimate for survival analysis is that it is univariable analysis which can only perform single factor analysis and it ignores the effects of other factors~\citep{Jager2008The}. The interaction between multiple factors has a great impact on survival analysis. Cox proportional hazards (CPH) model has the ability to perform multivariate survival analysis~\citep{D1972Regression} and is widely used in survival analysis research. Guo et al. studied the link between genetic variation and survival of breast cancer using the CPH model~\citep{guo2015identification}. With the advancing of high-dimensional molecular data or clinical data, the survival analysis model based on the CPH model has been gradually improved. Gui et al. proposed the LARS-Cox model which classified high-risk and low-risk patient groups using gene expression data relevant to the survival phenotypes~\citep{gui2005penalized}. Eric et al. proposed a framework to predict cancer subtypes using the combination of gene expression data and clinical data, and then proceeded with survival analysis of cancer~\citep{bair2004semisupervised}. Park et al. proposed an $L_{1}$-regularized CPH model which was applicable to survival analysis of data in which the number of variables was much larger than that of the samples~\citep{park2007l1regularization}. Tibshirani and Robert proposed a variable selection and compression method used for the CPH model~\citep{tibshirani1997the}. For high dimensional micro-array data, Houwelingen et al. added cross-validated partial likelihood into the CPH model to improve the survival prediction performance~\citep{vanhouwelingen2006crossvalidated}. To further improve the CPH model, Li et al. proposed the MTLSA model which used multi-task learning instead of the original survival analysis and made survival time prediction by estimating survival status at each survival time point~\citep{Li2016A}. For high-dimensional data, deep neural network (DNN) can also be used for survival analysis. In order to explore the nonlinear relationship between patient's covariates and treatment effectiveness, the DeepSurv model proposed by Katzman et al. replaced the loss function of the original DNN with the loss function in the CPH model.

Compared to molecular data or clinical data, pathological image data provides more direct and rich information about diseases. In the past few years, pathological-image based, especially WSI-based survival analysis has emerged with the development of pathological image technology. For non-small cell lung cancer (NSCLC) survival analysis, Wang et al. developed a framework which aimed to find image markers with strong survival prediction ability~\citep{wang2014novel}. Zhu et al. proposed a supervised conditional Gaussian graphical model (SuperCGGM) which combined pathological images and genetic data to make survival analysis~\citep{Zhu2016Imaging}. Yu et al. extracted 9,879 quantitative image features from the WSIs of lung adenocarcinoma and squamous cell carcinoma patients and selected top features using regularized machine leaning methods to classify shorter- and longer-term survivors~\citep{Yu2016Predicting}. Luo et al. proposed a pipeline which could automatically extract morphological features and make survival prediction of lung cancer patients~\citep{luo2017comprehensive}.

It is commonly known that deep learning has developed rapidly in recent years and has shown powerful image processing capabilities. Unlike hand-crafted feature extraction methods, deep learning methods can automatically extract high level features based on specific tasks and is capable of handling the situation when the data is quite high in volume. So far, deep learning has been applied to various image processing tasks and there are also many studies using deep leaning in survival analysis. For instance, Yao et al. extracted imaging bio-markers from histopathological images for lung cancer survival prediction based on deep CNNs~\citep{yao2016imaging}. Different from molecular profiling data and traditional imaging bio-markers, the imaging bio-makers also contained cell type information. Paul et al. extracted features from CT images of lung cancer using a pre-trained CNN and these features were used to make prediction of short and long term survival~\citep{paul2016combining}. Lao et al. fused deep imaging features with hand-crafted features from preoperative multi-modality MR images to make overall survival (OS) prediction of patients with glioblastoma multiforme (GBM)~\citep{lao2017a}. To analyze the rectal cancer survival, Li et al. proposed a framework to extract radiomic features using CNNs by optimizing partial likelihood of a proportional hazards model~\citep{Li2017Deep}. Zhu et al. proposed a WSISA framework to make lung cancer survival prediction based on WSIs using deep learning~\citep{zhu2017wsisa}. This study chose discriminative patches extracted from WSIs to formulate WSI-based survival analysis. Wang et al. innovatively studied the relationship between tumor shape and patient prognosis~\citep{wang2018comprehensive}. They extracted features which were related to patient survival outcome from tumor regions identified by CNN in lung cancer pathological images. Huang et al. proposed a survival model based on deep learning to predict patients survival outcomes using multi-view data which contained pathological images and molecular data~\citep{huang2018deep}.
%%remove mnih2014recurrent and mansimov2015generating and ba2014multiple after genenration
Besides, attention module is widely used in the field of image processing such as image classification and image generation~\citep{gregor2015draw,wang2017residual}. Attention module embedded in CNNs can help CNNs pay more attention to image region that contains the most useful information for the image processing task during the training process. While improving the feature extraction performance of CNNs, the attention module does not bring in a large number of parameters. Hu et al. proposed SENet, a CNN architecture with channel-wise attention module (CAM)~\citep{hu2019squeezeandexcitation}. Different from SENet which only focuses on channel-wise attention, Woo et al. proposed convolutional block attention module (CBAM)~\citep{woo2018cbam} with both channel-wise attention module (CAM) and spatial-wise attention module (SAM). In addition to CAM and SAM, here we also built neuron attention module (NAM) which was made up of multi-layer perceptron (MLP) and was embedded into the fully connected (FC) layers. In summary, our proposed DCAS model embedded CAM, SAM, and NAM into a survival convolutional neural network (SCNN) to further improve the performance of the image-based survival analysis.

It is necessary to carry out survival analysis for EOC patients since survival analysis can help doctors learn about patients' survival status and the disease trend over time, and to provide better treatment plan and medication. Based on the preoperative CT images, Lu et al. found 4 descriptors named radiomic prognostic vector (RPV) chosen among 657 quantitative mathematical descriptors to identify patients whose overall survival was less than 2 years~\citep{Lu2019A}. However, using both WSIs and deep learning to predict survival of EOC patients is still needing more exploration.

\section{Methods}
As far as we know, we are the first to exploit the EOC survival prediction based on WSI and deep learning simultaneously. In this section, we will introduce our proposed EOCSA framework in details. The pipeline of the EOCSA framework is shown in Fig.~\ref{EOCSA}. We first randomly extracted patches from EOC patients' WSIs and clustered these extracted patches by a K-means algorithm. Next, we trained our proposed DCAS model for each cluster and extracted patch-level features of the selected clusters. At last, we generated patient-level features based on patch-level features to make final survival analysis using the LASSO-Cox model.

\subsection{Dataset}
The Cancer Genome Atlas (TCGA) collects cancer samples from over 11,000 patients~\citep{weinstein2013the}. It includes not only molecular profiling data but also high-quality pathological image data~\citep{kandoth2013mutational}. We downloaded the WSIs of EOC patients' and clinical data containing overall time and survival status from TCGA-OV~\citep{bell2011integrated}. There are 583 patients and 1,296 WSIs being selected from the TCGA-OV project in our study. Of these patients, 349 have died during follow-up studies and 234 are still alive at the end of the follow-up studies.

\subsection{Data Pre-processing}

\subsubsection{Extracting patches from WSIs}
Due to the high resolution nature of WSIs, it is time-consuming to process the whole image at one time. Some studies extracted patches from annotated region of interests (ROIs)~\citep{zhu2016deep}. The shortcomings of ROI-based methods are obvious. Firstly, ROI requires the participation of pathologists and it is laborious to annotate such large WSIs; Secondly, focusing on ROI ignores the impact of other regions. Other studies densely tiled the input images with non-overlapping windows~\citep{coudray2018classification}. The disadvantage of this method is that it increases the cost of computation due to the processing of some patches that do not contribute much.

We intended to extract patches automatically without the involvement of pathologists. Meanwhile, to reduce the computation cost, we used patches (with size of 512$\times$512) randomly extracted from WSIs to represent the whole images. Here we set a sampling ratio to control the number of the extracted patches. This number was calculated as follows:
\begin{equation}
\label{num}
Num = \frac{h \cdot w}{s^{2}}\times ratio
\end{equation}
where $h$ and $w$ denote the height and width of WSIs respectively; $s$ means the side length of each patch; The $ratio$ represents the sampling ratio and is set to 0.05 by experience.

Since the background might cover large area of the extracted patches, we used the following selection strategy to ensure that every extracted patch contains enough pathological areas. Firstly, we converted the patch from RGB space to gray space. It was proved by practice that pixel with a value larger than 200 belonged to the background area. So we set a threshold to control the proportion of background area in patches. In the specific implementation, we set this threshold to 0.5. During the image extraction process, we discarded the patch which contained more than half of the background area of the overall patch and re-extracted a patch with less background. Finally, for each WSI, we extracted $Num$ patches that contained most of the foreground which $Num$ was calculated by Eq.\ref{num}. 

\subsubsection{Clustering by K-means}
Both tumor regions and non-tumor regions are included in the WSIs. So the patches might be extracted from the tumor region, non-tumor region, or both. Unsupervised clustering was used to separate these patches in order to fully utilize the patches containing as many tumor regions. Following~\citep{zhu2017wsisa}, we first resized the patches to a much smaller size (50$\times$50); Then we flattened the patches and obtained 2,500 features for each patch; Next, we reduced the dimension of the features (here we reduced it to 50) based on principal component analysis (PCA) to reduce the computation cost; At last, the K-means algorithm was adopted to cluster the patches based on the image features. K-means minimizes the following cost function:
\begin{equation}
cost = \sum_{i=1}^{p}\sum_{n_{j}\in C_{i}}\left \| n_{j}-\mu _{i}\right \|^{2}
\end{equation}
where $ p $ means the number of clusters; $ n_{j} $ means the $ j $-th patch; $ C_{i} $ represents the $ i $-th cluster; and $ \mu_{i} $ indicates the center of the $ i $-th cluster.

\subsection{Feature Extraction}

\subsubsection{DCAS Model}
\begin{figure*}
	\centering
	\includegraphics[width=4.7in]{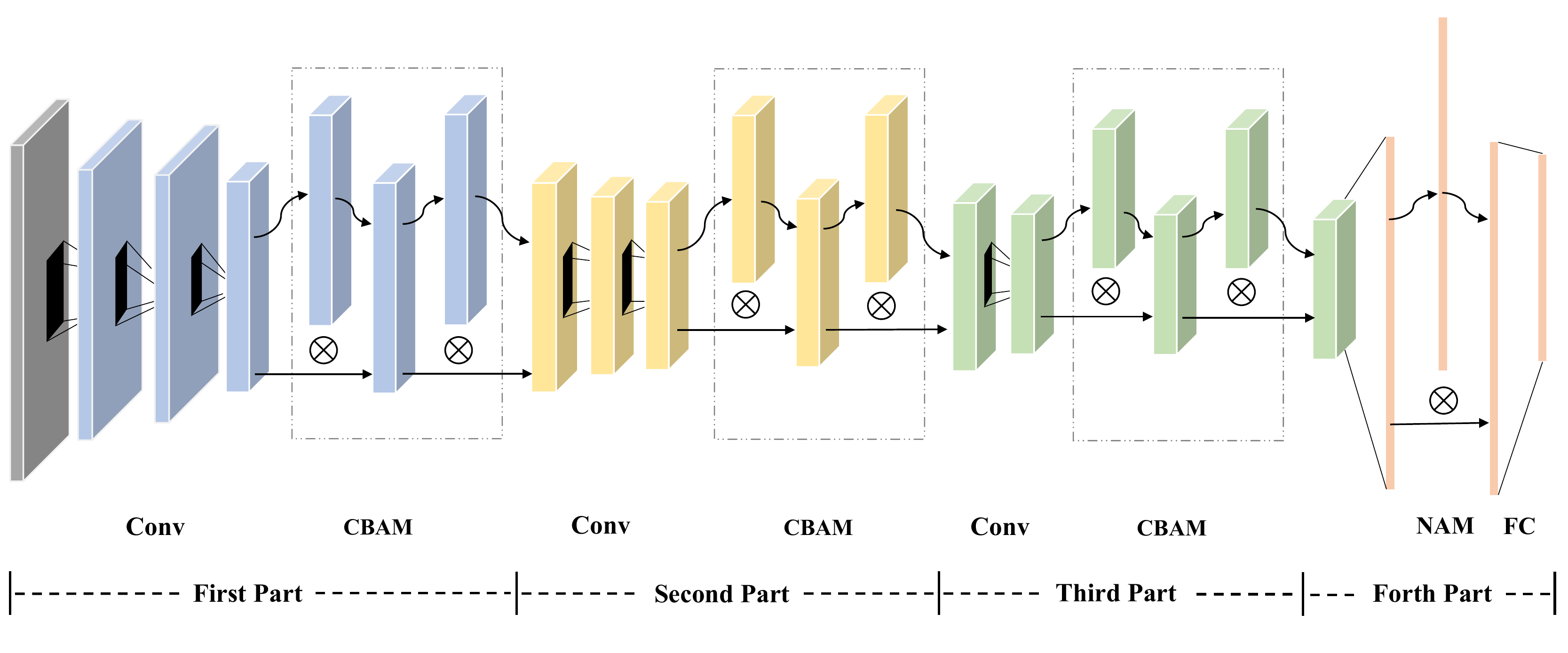}
	\caption{The structure of the DeepConvAttentionSurv (DCAS) model. The DCAS model has four parts. The first part consists of three Convolutional (Conv) layers, a Convolutional Block Attention Module (CBAM), a Batch Normalization (BN) layer and a MaxPooling layer; The second part consists of two Conv layers and a CBAM; The third part consists of a Conv layer, a CBAM, a BN layer and a MaxPooling layer; The fourth part consists of a Neuron Attention Module (NAM) and a Fully Connected (FC) layer. We defined the loss function as negative log partial likelihood so that DCAS can perform survival analysis tasks. $ \otimes $ represents element-wise multiplication.}
	\label{DCAS}
\end{figure*}

The DCAS model is a deep network model we proposed here for survival analysis. It is made up of four parts. The first part consists of three convolutional (Conv) layers, a CBAM, a batch normalization (BN) layer and a MaxPooling layer. The second part consists of two Conv layers and a CBAM; The third part consists of a Conv layer, a CBAM, a BN layer and a MaxPooling layer. The last part consists of a NAM and a FC layer. The architecture of the DCAS model is shown in Fig.~\ref{DCAS} and the parameter setting is shown in TABLE~\ref{p_setting}.

\begin{table}[t]
	\centering
	\caption{The parameter setting of DeepConvAttentionSurv (DCAS).}
	\label{p_setting}
	\begin{tabular}{cccccc}
		\toprule
		& Layer& Kernel& Channel& Stride& Reduce\\
		\midrule
		{\multirow{6}{*}{First part}}
		& Conv1\_1 & 3*3 & 32& 4& -\\
		& Conv1\_2 & 3*3 & 32& 1& -\\
		& Conv1\_3 & 3*3 & 32& 1& -\\
		& CBAM & -& -& - & 4\\
		& BN & -& -& - &- \\
		& MaxPooling & 2*2& -& -& -\\
		\hline
		{\multirow{3}{*}{Second part}}
		& Conv2\_1 & 3*3 & 32& 2& -\\
		& Conv2\_2 & 3*3 & 32& 1& -\\
		& CBAM & -& -& -& 4\\
		\hline
		{\multirow{4}{*}{Third part}}
		& Conv3 & 3*3 & 32& 2& -\\
		& CBAM & -& -& -& 4\\
		& BN & -& -& -& -\\
		& MaxPooling & 2*2& -& -& -\\
		\hline
		{\multirow{2}{*}{Fourth part}}
		& NAM & -& -& -& 32\\
		& FC & -& 32& & -\\	
		\bottomrule	
	\end{tabular}
\end{table}

The CBAM~\citep{woo2018cbam} contains channel attention module (CAM) and spatial attention module (SAM), focusing on image channel and spatial relations respectively. Unlike the previous attention module, which only focuses on `where' or only `what', the CBAM focuses on both `where' and `what'. In CAM, both average-pooled features and max-pooled features were used to better represent the input feature maps. With negligible overheads, embedding CBAM can help DCAS pay more attention to the tumor area of the input images to extract high quality features. The output of CBAM is calculated as follows:
\begin{equation}
CBAM = O_{SAM}(O_{CAM}(In)\otimes In)\otimes (O_{CAM}(In)\otimes In)
\end{equation}

\begin{equation}
\label{cam}
O_{CAM}=\delta(MLP(AP(In))+MLP(MP(In)))
\end{equation}

\begin{equation}
\label{sam}
O_{SAM}=\delta(Conv_{7\times7}(Concat(AP(In_{SAM}),MP(In_{SAM}))))
\end{equation}
where $ In $ indicates input feature maps of CBAM. $ O_{CAM} $ and $ O_{SAM} $  mean channel attention operation and spatial attention operation which are shown in Eq.~\ref{cam} and Eq.~\ref{sam} respectively. $ \otimes $ represents element-wise multiplication. $ \delta $ denotes the sigmoid function. $AP$ and $MP$ mean average pooling and max pooling respectively. $Conv_{7\times7}$ means convolution operation with $7\times7$ filter size. $In_{SAM}$ represents the input of $O_{SAM}$ and it is also the output of $O_{CAM}$.

We added NAM before the last FC layer to emphasize the features which had more contributions. The NAM is composed of MLP which contains one hidden layer. The hidden activation size is calculated by $ Num/Reduce $, where $ Num $ means the input size of the NAM and $ Reduce $ is shown in TABLE~\ref{p_setting}. The output of NAM was calculated as follows:
\begin{equation}
NAM =  \delta (MLP(Input)) \otimes Input
\end{equation}
where $ \delta $ denotes the sigmoid function. Through the sigmoid function, we obtained the weight coefficient ranging from $ 0 $ to $ 1 $.

Similar to DeepConvSurv~\citep{zhu2016deep}, the loss function of the proposed model was defined the same as the Cox model's loss function~\citep{D1972Regression}, which was represented as negative log partial likelihood:
\begin{equation}
Loss = -\sum_{i:R_{t_{i}}=1}\left (Risk(x_{i})-log\sum_{j:t_{j}>=t_{i}}e^{Risk(x_{j})}\right )
\end{equation}
where $ R_{t_{i}} $ is the risk set at time $ t_{i} $. It is a set which contains the patients who are still under follow-up studies before $ t_{i} $. $x_{i}$ and $x_{j}$ represent the $i^{th}$ and $j^{th}$ inputs. $Risk(x_{i})$ means the output of the DCAS model for the input $x_{i}$.

\subsubsection{Cluster Selection and Feature Extraction}\label{sec:clusterselection}
Obviously, patches contribute differently to the performance of survival analysis. As mentioned before, K-means was able to group the similar patches together. Based on the clustering results, we picked the clusters which were more informative and predictive. We trained the DCAS model for each cluster and selected the clusters whose predictive accuracy was higher than a preset threshold. Then we extracted the deep features for all the patches in the selected cluster from the last FC layer of the corresponding trained DCAS model to generate patch-level features.

\subsection{Final Survival Analysis}

\subsubsection{Weighted Patient-level Features}
We generated weighted patient-level features based on the extracted patch-level features. From the above description, the patches extracted from one patient's WSI might be assigned to different clusters due to different predictive ability of the patches. As described in Section~\ref{sec:clusterselection}, informative clusters were selected through their performance based on the DCAS model. If one patient had more clusters being selected, it means that this patient's features might be more predictive. Thus we set a larger weight to the patient-level feature with more clusters being selected. The weight was calculated as in Eq.~(\ref{eq:weight}).
\begin{equation}\label{eq:weight}
w_{i} = \frac{n_{i}}{n_{h}-n_{l}},\qquad i\in\{1,2,\cdots,N\}
\end{equation}
where $ n_{i} $ denotes the number of clusters covered by the $ i $-th patient's patches; $ n_{h} $ and $ n_{l} $ mean the maximal and minimal numbers of clusters among all the patients; $ N $ represents the number of patients. With the weight of each patient, the weighted patient-level features can be calculate by Eq.~(\ref{eq:weightfeature}):
\begin{equation}\label{eq:weightfeature}
F_{i} = w_{i}*(\sum_{j=1}^{C_{i}}( \sum_{m=1}^{L_{ij}}f_{m}/L_{ij})/C_{i})
\end{equation}
where $ C_{i} $ is the number of clusters covered by patches from patient $ i $; $ L_{ij} $ is the number of patches extracted from patient $ i $ in the cluster $j$. $ f_{m} $ means the patch features extracted from the patch which is in cluster $ m $ by the DCAS model. Specifically, in order to obtain the weighted features of a patient, we first searched the clusters containing patches from this patient. Then, for each cluster, we averaged the patch-level features extracted by the DCAS model to obtain the cluster-based patient-level features. After that, we averaged the cluster-based patient-level features from all clusters which contained the patches of this patient to obtain the patient-level features. Finally, we weighted the patient-level features through Eq.~(\ref{eq:weight}).

%where $ L_{i} $ is the number of patches extracted from patient $ i $; $ I_{i} $ is the number of WSIs of patient $ i $ because in the TCGA-OV dataset, a patient may have one or more WSIs; $ C_{i} $ is the number of clusters covered by patches from patient $ i $; $ f $ means the features in patches extracted by the trained DCAS.

\subsubsection{Final Survival Analysis}
We made the final survival analysis based on the weighted patient-level features and the LASSO-Cox model~\citep{tibshirani1997the}. We used 10-fold cross validation to obtain a reliable and stable model. The algorithm of proposed EOCSA is described in Algorithm~\ref{eocsa_algorithm}.

\begin{algorithm}[t]
	\setstretch{1}	
	\caption{EOCSA Algorithm}
	\label{eocsa_algorithm}
	\hspace*{0.02in} {\bf Input:}
	WSIs, height of WSIs \textbf{\emph{h}}, width of WSIs \textbf{\emph{w}}, survival time \textbf{\emph{t}}, survival status \textbf{\emph{ss}}, sample ratio \textbf{\emph{ratio}}, side length of patches \textbf{\emph{s}}, thumbnail size \textbf{\emph{ts}}. \\
	\hspace*{0.02in} {\bf Output:}
	The risk of patients
	\begin{algorithmic}[1]
	  \STATE ** The fisrt step: \textbf{Data Pre-processing} **
	  \FOR{all WSIs}
	  \STATE $num\_patches=\frac{h \cdot w}{s^{2}}\times ratio$
	  \ENDFOR
	  \FOR{all patches}
	  \STATE thumbnails = createThumbnail(patches,ts)
	  \STATE pca = PCA(thumbnails)
	  \STATE clusters = K\_means(pca,num\_clusters)
	  \ENDFOR
	  \STATE ** The second step: \textbf{Feature Extraction} **
	  \FOR{cluster in clusters}
	  \STATE model = trainDCAS(cluster\_train,\emph{t},\emph{ss})
	  \STATE test\_CI = evaluate(cluster\_test)
	  \ENDFOR
	  \STATE selected\_clusters = selectCluster(test\_CI)
	  \STATE extracted\_features = trainedDCAS(selected\_clusters)
	  \STATE ** The third step: \textbf{Final Analysis} **
	  \STATE weighted\_patient-level\_features = weightedFeature(extracted\_features)
	  \STATE risk = LASSO-Cox(weighted\_patient-level\_features,t,ss)
	  \RETURN The risk of patients
    \end{algorithmic}
	
\end{algorithm}

\subsection{Performance Estimation}
We evaluated the performance of the proposed method using concordance index (C-index). C-index quantifies the goodness of fit in a logistic regression model and the value ranges from 0 to 1. A larger value of C-index indicates better discriminative ability of the model while a smaller value shows poorer discriminative ability. C-index was calculated as follows:
\begin{equation}
CI = \frac{1}{n} \sum_{i\in\{1\cdots N\}}\sum_{s_{j}>s_{i}}I[Risk(x_j)>Risk(x_i)]
\end{equation}
where $ n $ is the number of valid pairs. $ I[\cdot] $ is the indicator function. $ s_{[\cdot]} $ means the actual observation time. $x_{i}$ and $x_{j}$ represent the $i^{th}$ and $j^{th}$ inputs. $Risk(x_{i})$ means the output of the DCAS model for the input $x_{i}$ and the same as $Risk(x_{j})$. In addition, we used the receive operating characteristic curve~(ROC curve) and the area under the ROC (AUC) to evaluate the performance.

\section{Experimental Results}

\subsection{Hyperparameter settings of EOCSA}
In this study, we proposed a EOCSA framework for EOC survival analysis. We extracted the patch-level features using our proposed DCAS model and generated patient-level features with the proposed weight calculation method to carry out survival analysis of EOC. Three steps were involved in the EOCSA framework.

In the first step, we randomly extracted patches from the WSIs with 20$ \times $ magnification (0.5 microns per pixel). We removed the patches with little foreground, that is, the patches with quite high percentage of the background region (values of pixels exceed 200 in the gray space converted by RGB space)~\citep{Yu2016Predicting}. Then the patches were re-sized to 512*512 and were clustered based on K-means algorithm. The obtained numbers of patients, WSIs, and patches are shown in TABLE~\ref{data_description}. The sampling ratio was set to 0.05. We clustered those patches to 10 groups by K-means algorithm.

\begin{table}[]\normalsize
	\centering
	\caption{The numbers of patients, WSIs, and patches.}
	\label{data_description}
	\begin{tabular}{cc}
		\toprule
		Dataset&TCGA-OV\\
		\midrule
		Number of patients&583\\
		Number of WSIs&1,296\\
		Number of patches&175,712\\
		\bottomrule	
	\end{tabular}
\end{table}

In the second step, we trained the DCAS model for each cluster, selected clusters with more predictive ability, and extracted patch-level features. In the DCAS model, we added a BN layer to reduce over-fitting and used ReLU as the activation function. We used 80\% of the patches in each cluster as training set and the remaining 20\% as test set. Each DCAS model was trained with 150 epochs. The batch size was set to 256. Stochastic gradient descent (SGD) optimizer was used with a learning rate at 0.1. During training procedure, the learning rate was reduced by half every 20 epochs. Finally, we selected the cluster with test predictive accuracy higher than 55.0\%. We took the last FC layer as the feature extraction layer and extracted 32 features for each patch.

In the third step, we obtained patient-level features using the proposed weight calculation method and carried out survival analysis using the LASSO-Cox model. For obtaining stable and reliable model, we used 10-fold cross validation in the survival analysis process.

\subsection{Cluster Selection and Patch-level Feature Extraction}
\begin{figure}
	\centering
	\includegraphics[width=4in]{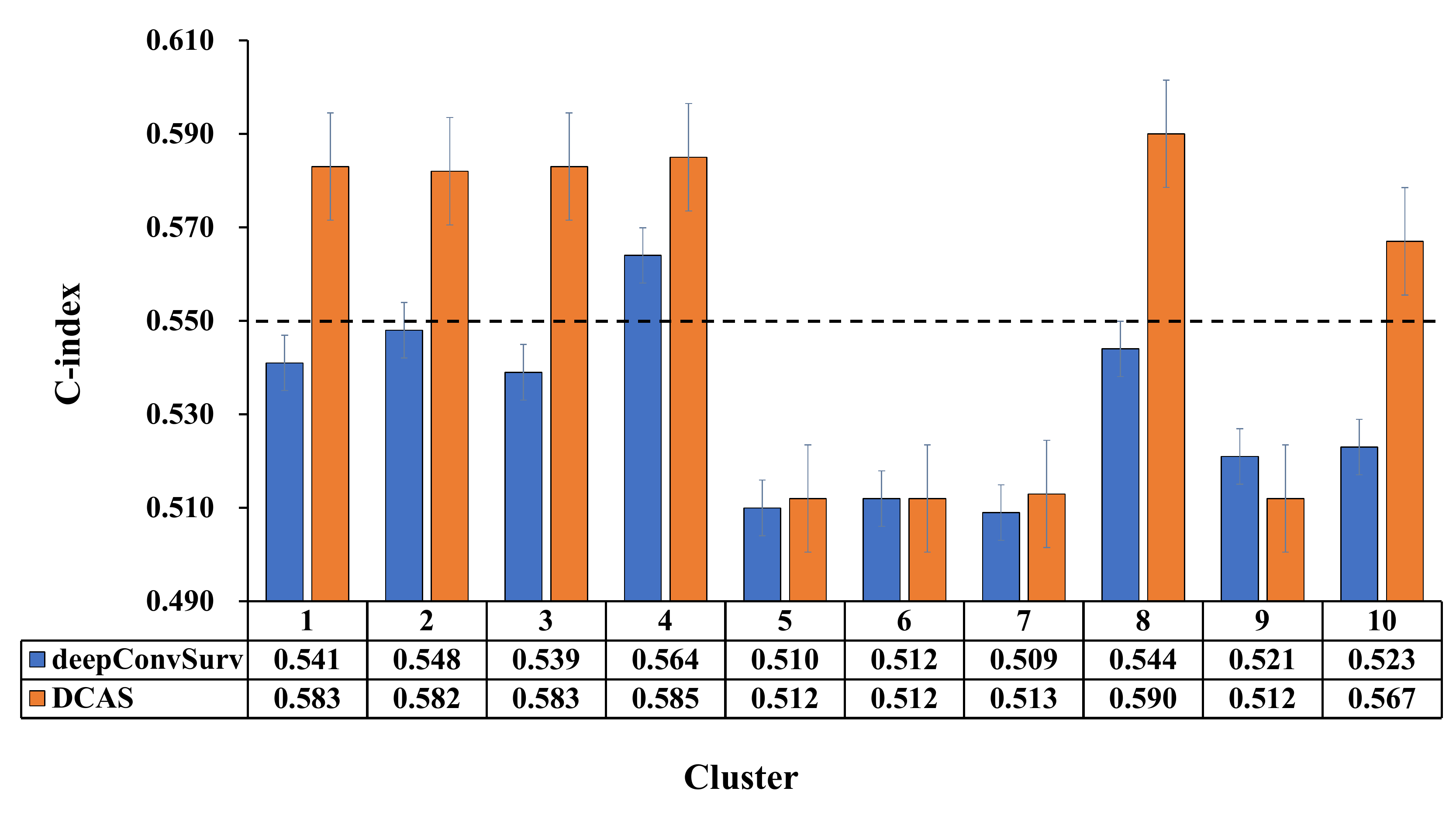}
	\caption{The comparative results between deepConvSurv and DeepConvAttentionSurv (DCAS) in each cluster. The horizontal dotted line shows the cluster selection threshold we have set. The C-index values of six clusters stay above the threshold value in DCAS while only one of deepConvSurv is higher than the threshold.}
	\label{modelComparison}
\end{figure}

The deepConvSurv model is the first CNN-based survival prediction model based on WSIs. In our experiment, the model was trained within each cluster to extract patch-level features. The prediction results of the deepConvSurv model and the DCAS model are shown in Fig.~\ref{modelComparison}.

In Fig.~\ref{modelComparison}, compared to the deepConvSurv model, nine of the ten clusters show higher or equal C-index values for the DCAS models. The overall C-index for DCAS is 0.228 higher than that of deepConvSurv. It is known that a higher C-index value indicates higher quality of the features extracted by the trained models. Our proposed DCAS model with attention modules including CAM, SAM, and NAM makes use of the most informative patches instead of the whole image slides so more predictive model and higher-quality features are obtained to achieve more accurate results.

Furthermore, in~\citep{zhu2017wsisa}, the threshold for the C-index was set to 0.5, which resulted in that all clusters trained by deepConvSurv and DCAS based on WSIs were chosen. Thus the less predictive clusters were not eliminated and redundant image information was mixed. In our study, we set the threshold for the C-index to 0.55 (shown as the dotted line in Fig.~\ref{modelComparison}). Six clusters were chosen in DCAS and only one cluster was selected in deepConvSurv.

\subsection{Survival Model Determination in EOCSA}
The survival model predicts the survival time based on the generated patient-level features, so the performance of the survival model is also quite important in the whole framework. Here we compared the performance of five most advanced survival models, $ l_{1} $-norm Cox model (LASSO-Cox)~\citep{tibshirani1997the}, elastic-net penalized Cox model (EN-Cox)~\citep{Yi2013A}, random survival forests (RSF)~\citep{ishwaran2008random},  boosting concordance index (BoostCI)~\citep{mayr2014boosting} and multi-task learning model for survival analysis (MTLSA)~\citep{Li2016A} and sought to determine the most proper survival model.

LASSO-Cox simultaneously performs variable selection and shrinkage in CPH. In EN-Cox, different from LASSO-Cox, elastic-net (EN) is used for variable selection and shrinkage in the CPH. RSF builds an integrated model based on decision tree, which greatly improves the prediction performance. BoostCI is a non-parametric measure to evaluate the prediction ability based on the concordance index. MTLSA makes survival analysis by reformulating the survival analysis problem to a multi-task learning problem.

\begin{figure}
	\centering
	\includegraphics[width=4in]{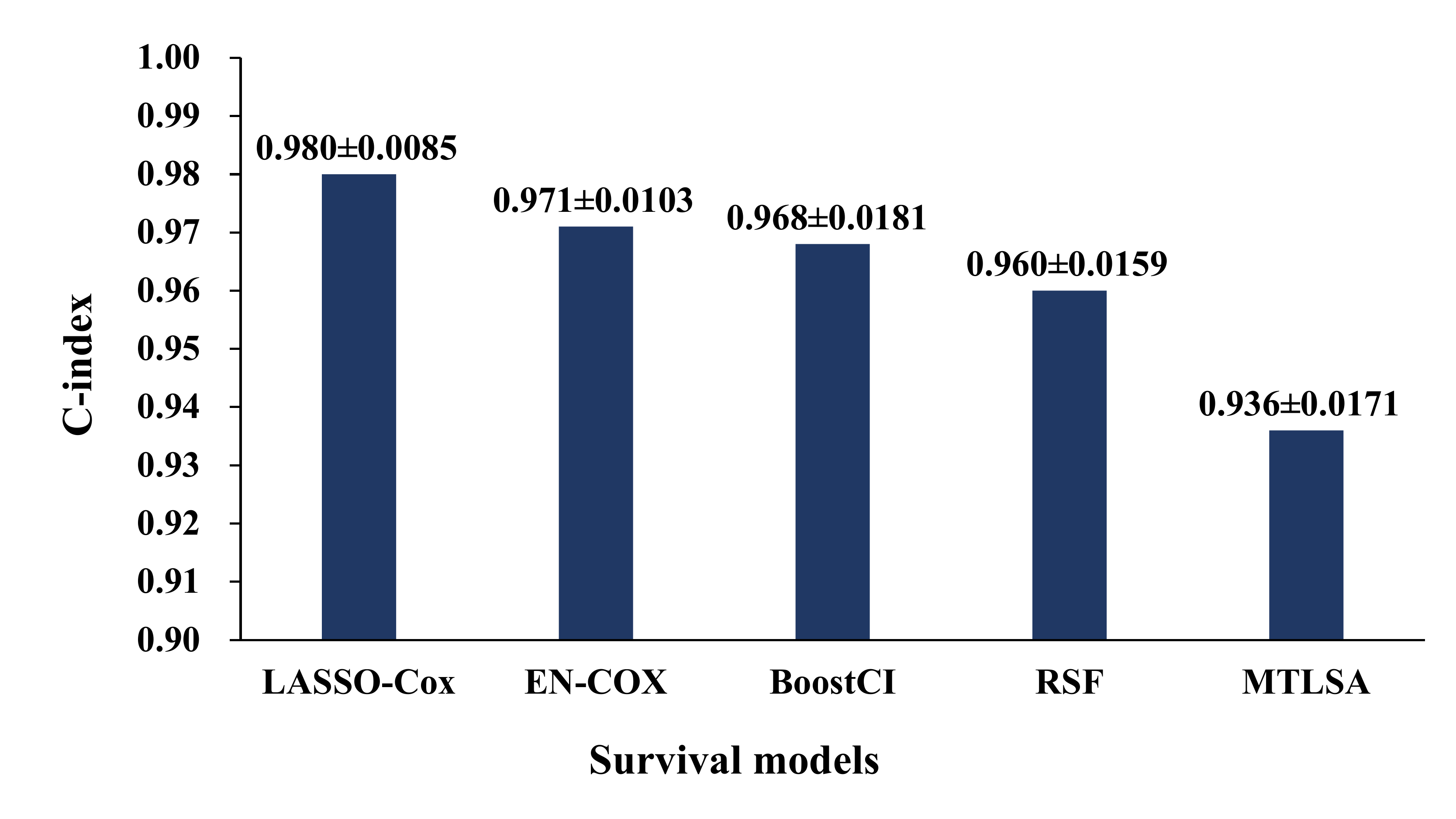}
	\caption{The results of different survival models in EOCSA framework. The LASSO-Cox model achieves the best prediction performance.}
	\label{modelsInEOCSA}
\end{figure}

Fig.~\ref{modelsInEOCSA} shows the survival analysis results of the five survival models. LASSO-Cox model achieves the highest C-index value of $ 0.980 $. The C-index values of En-Cox, BoostCI, RSF, and MTLSA are $ 0.971 $, $ 0.968 $, $ 0.960 $, and $ 0.936 $ respectively. Therefore, LASSO-COX was chosen to perform the final survival estimates.

\subsection{Comparison with WSISA}

\subsubsection{Comparison of The Patient-level Features}
\begin{figure}
	\centering
	\includegraphics[width=4in]{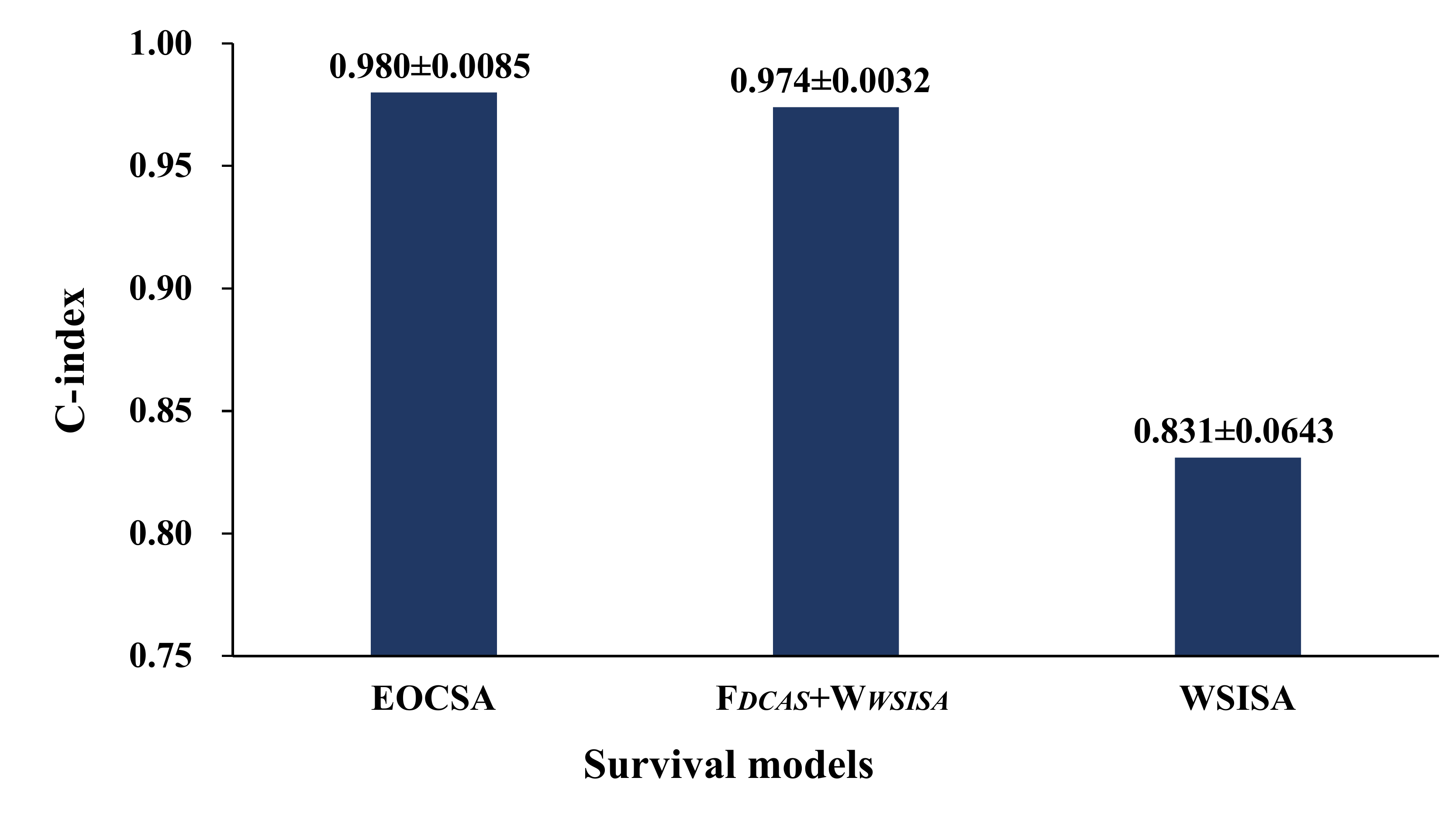}
	\caption{The comparison of features used in different frameworks. F$_{DCAS}+$W$_{WSISA}$ means the patient-level features extracted by DCAS models with WSISA's weights.}
	\label{weighted_comparison}
\end{figure}

Compared to the patient-level features used in the WSISA framework, features were weighted differently in the EOCSA according to their contributions. For fair comparison, we compared three feature sets: features used in EOCSA, features generated based on DCAS (denoted with F$_{DCAS}$) but with WSISA's weights (denoted with W$_{WSISA}$), and features used in WSISA. We show the results in Fig.~\ref{weighted_comparison}, which shows that the EOCSA features achieves the highest and the WSISA has the lowest C-index value, and the C-index value of EOCSA is $0.006$ higher than that of F$_{DCAS}+$W$_{WSISA}$. This indicates the effectiveness of the features extracted by DCAS and the weight in EOCSA is more accurate than that in WSISA.

\subsubsection{Survival Analysis Comparison}
\begin{figure}[t]
	\centering
	\subfigure[One year]{
		\includegraphics[width=5.5cm]{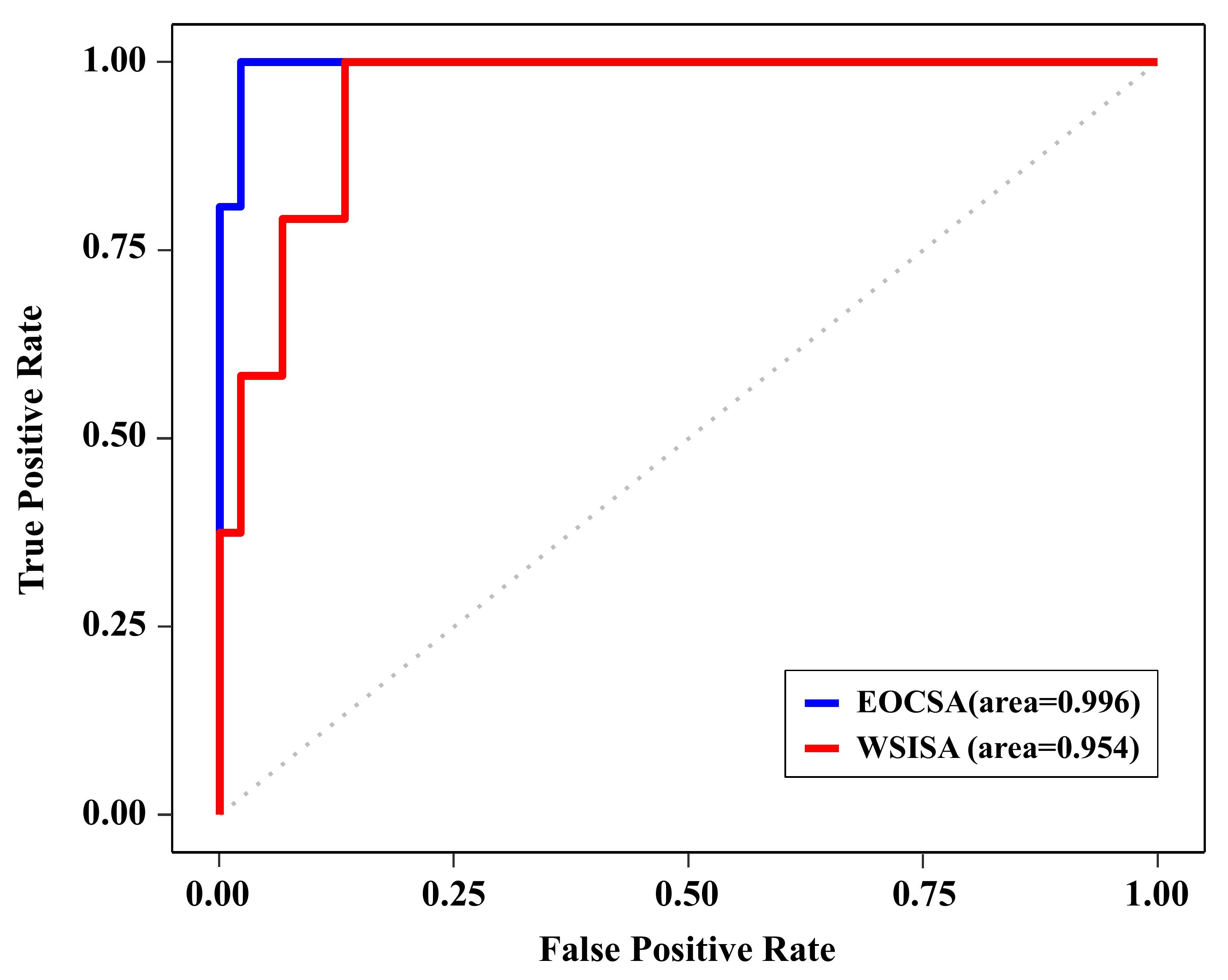}
	}
	\subfigure[Three years]{
		\includegraphics[width=5.5cm]{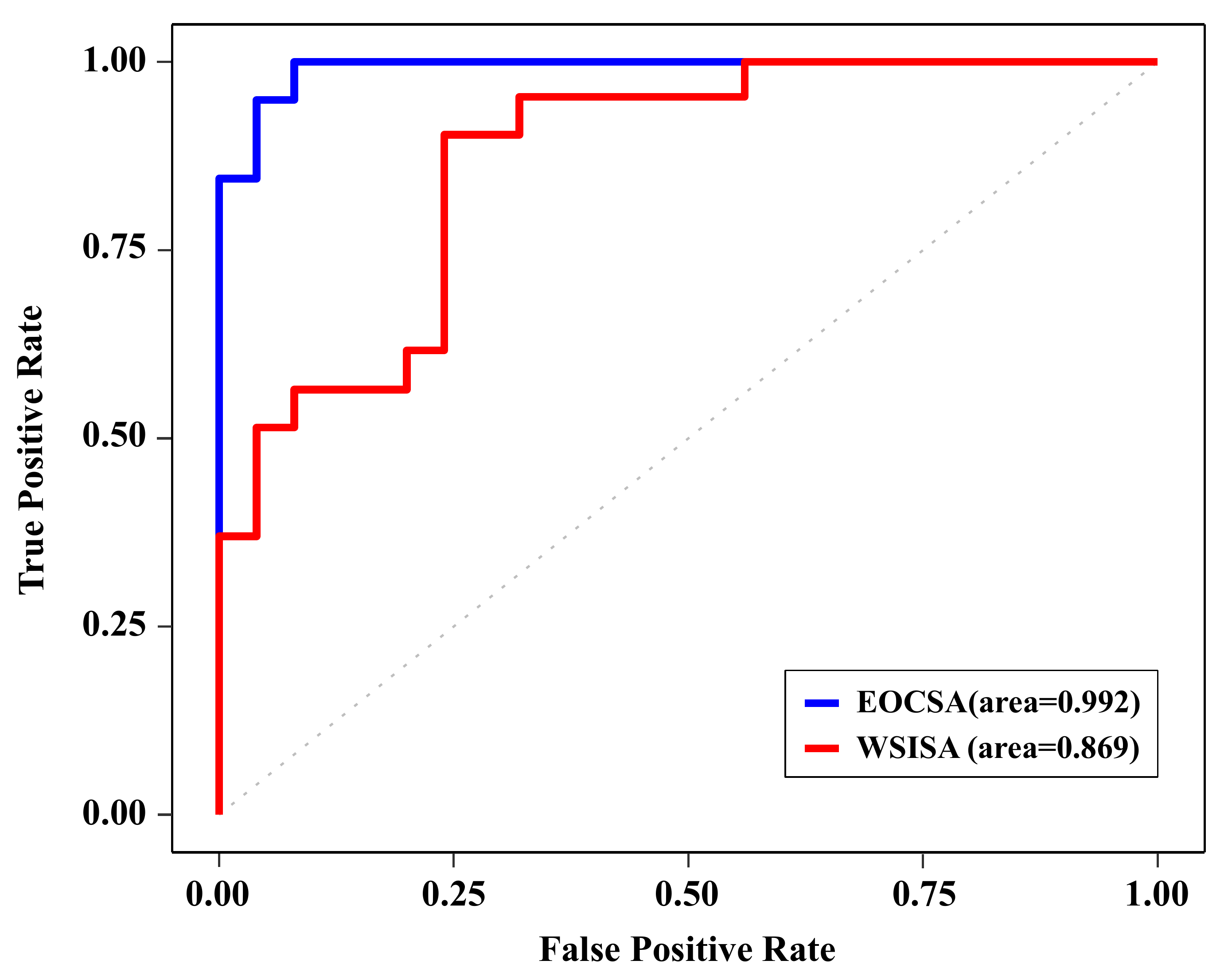}
	}
	\subfigure[Five years]{
		\includegraphics[width=5.5cm]{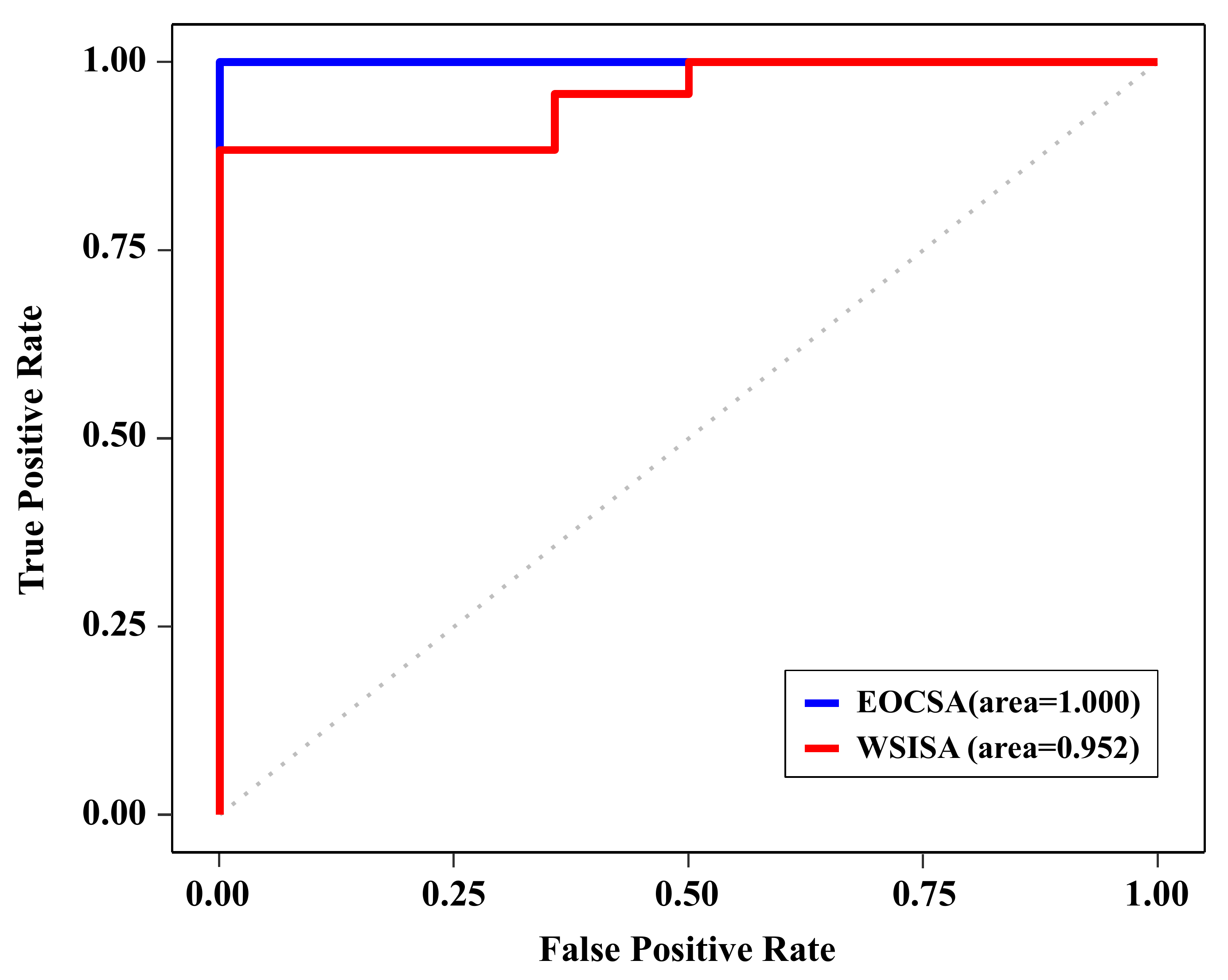}
	}
	
	\caption{The comparison of EOCSA and WSISA survival ROCs when setting with different prediction time.}
	\label{Compare_ROC}
\end{figure}

\begin{figure}[t]
	\centering
	\begin{subfigure}[EOCSA]{
			\label{survival_EOCSA_WSISA:a}
			\includegraphics[width=5.8cm]{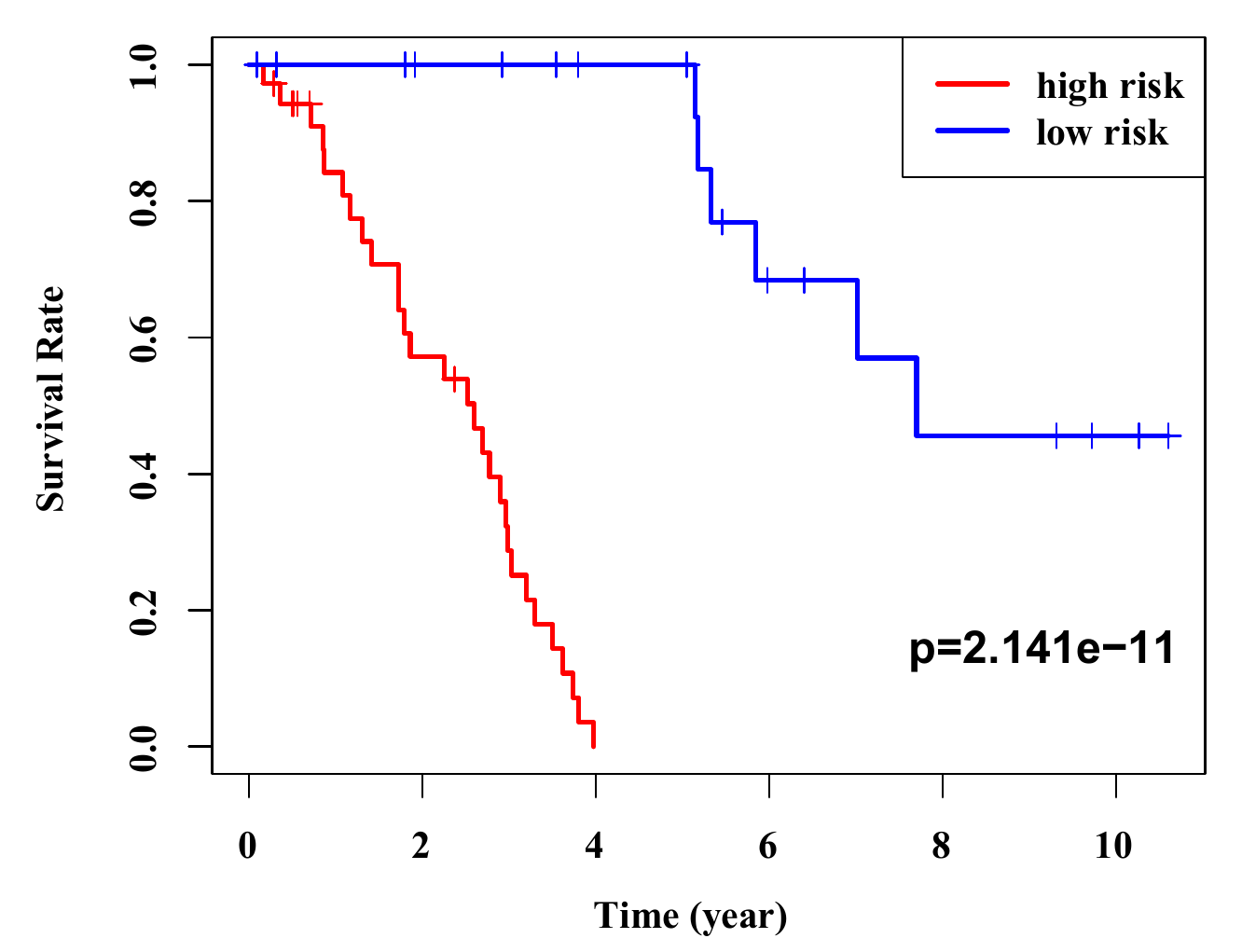}}
	\end{subfigure}
	\begin{subfigure}[WSISA]{
			\label{survival_EOCSA_WSISA:b}
			\includegraphics[width=5.8cm]{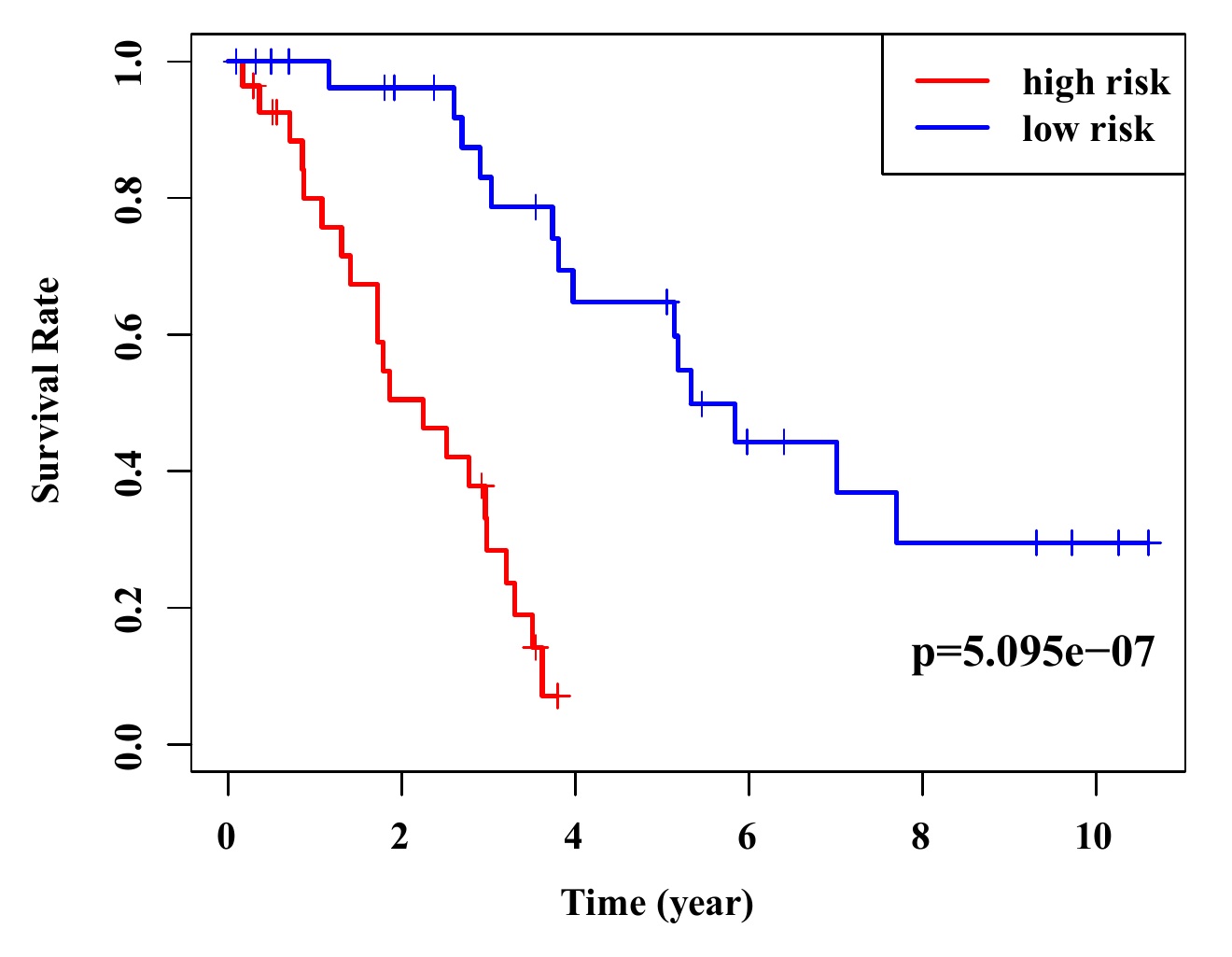}}
	\end{subfigure}
	\caption{The ROCs of EOCSA and WSISA for high risk and low risk prediction.}
	\label{survival_EOCSA_WSISA}
\end{figure}

We first compared ROC curves between the EOCSA framework and the WSISA framework. The comparison result is shown in Fig.~\ref{Compare_ROC}. The AUC values of EOCSA are 0.996, 0.992 and 1 with predict time being set to 1, 3 and 5 years respectively. And the AUC values of EOCSA are $0.042$, $0.123$, $0.048$ higher than those of WSISA when the predict time are 1, 3 and 5 years respectively. We also compared the prediction of high-risk/low-risk patients predicted by EOCSA and WSISA and the survival curves can be found in Fig.~\ref{survival_EOCSA_WSISA:a} and Fig.~\ref{survival_EOCSA_WSISA:b} respectively. We set patients with a risk score higher than the median risk score as high-risk patients, and patients with a risk score lower than the median risk score as low-risk patients. From the figures, the p-value of our proposed EOCSA is much smaller than that of WSISA, which demonstrates the prediction for high risk/low risk of EOCSA is more reliable than WSISA.

In our experiments, we chose the results of one fold for comparison. The output risk score of the LASSO-Cox model was used to generate the time-dependent survival ROC. The predict time was set to 1, 3 and 5 years.

\begin{comment}
\begin{figure}
\centering
\includegraphics[width=2.5in]{figs/survival_EOCSA.pdf}
\caption{The ROCs of EOCSA for high risk and low risk prediction.}
\label{survivalEOCSA}
\end{figure}

\begin{figure}
\centering
\includegraphics[width=2.5in]{figs/survival_WSISA.pdf}
\caption{The ROCs of WSISA for high risk and low risk prediction.}
\label{survivalWSISA}
\end{figure}
\end{comment}

\subsection{Comparison between Phenotype Features and Genotype Features}
We also compared the performance using phenotype features and genotype features. To use the genotype features, similar to EOCSA, we established a gene expression level-based framework to predict the survival for EOC patients. Firstly, we built a deepSurv model which was constructed in~\citep{katzman2018deepsurv} for predicting survival based on genetic data. The structure of the deepSurv contains four FC layers and the number of neurons in each FC layer was set to $ 1024 $, $ 128 $, $ 64 $ and $ 32 $. The last FC layer was set as a feature extraction layer. We added BN layers in the deepSurv to avoid over-fitting. In order to generate patient-level features, we randomly divided the patients into five groups and obtained the features for each patient. Then we repeated this procedure for five times. The final patient-level features were calculated through averaging the corresponding feature values of the five repeats. At last, we made the final survival prediction based on the patient-level genetic features and the LASSO-Cox model.

\begin{figure}
	\centering
	\includegraphics[width=4in]{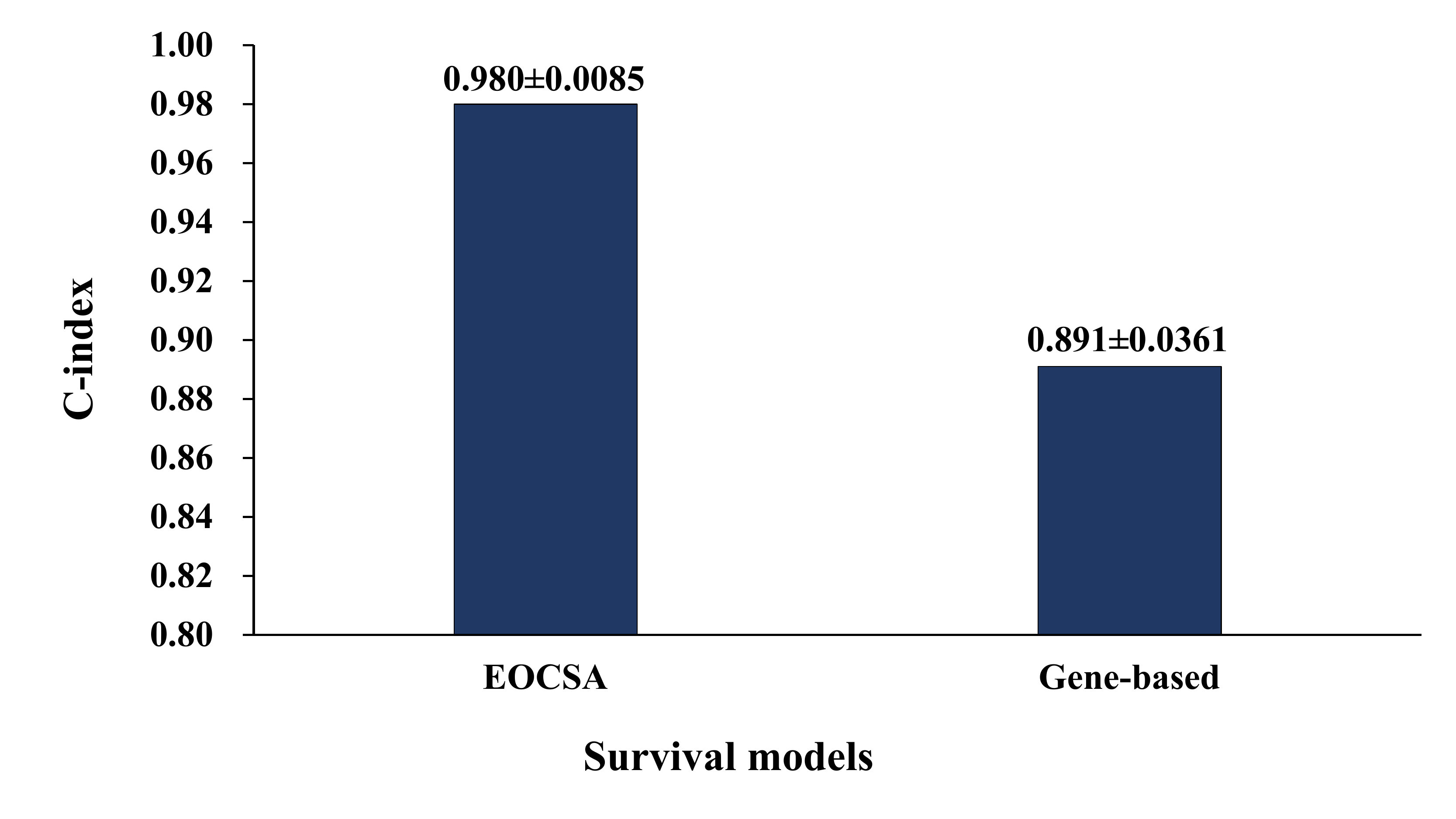}
	\caption{The results of the EOCSA framework and the gene expression level-based framework. Our proposed EOCSA is 0.089 higher than the gene-based framework on the C-index values.}
	\label{EOCSA_GENE}
\end{figure}

Since only 373 patients in TCGA-OV have complete genetic data, the WSIs of the same 373 patients were picked for fair comparison. The comparison result is shown in Fig.~\ref{EOCSA_GENE}. It shows that the C-index values of EOCSA and the gene-based model are $ 0.980 $ and $ 0.891 $ respectively. The C-index value for the EOCSA framework is $ 0.089 $ higher than that of the gene-based framework. This indicates that the phenotype features are more discriminative than the genotype features.

\subsection{Comparison Between EOCSA and Other Survival Models}
Currently, most of the image-based survival analysis methods extracted predefined hand-crafted features (denoted with ``F$_{hand}$'') and then conducted prediction based on algorithms such as LASSO-COX or En-Cox. Here we compared the performance between the EOCSA and other survival frameworks including WSISA, F$_{hand}$+LASSO-COX, F$_{hand}$+EN-COX, F$_{hand}$+RSF, F$_{hand}$+BoostCI, and F$_{hand}$+MTLSA. We firstly extracted hand-crafted features for each patch (sized with $ 512*512 $). Based on~\citep{zhu2017wsisa, Yu2016Predicting}, we extracted 562 quantitative features from each patch. These features included size, shape, and texture of the objects including cell, nuclei, and cytoplasm, as well as relationship between neighbouring objects (the details can be found in the supplementary materials). And then we generated patient-level features by averaging all the patches' features of the same patient. At last, the patient-level features were employed to carry out the final survival analysis.

\begin{figure}
	\centering
	\includegraphics[width=4.5in]{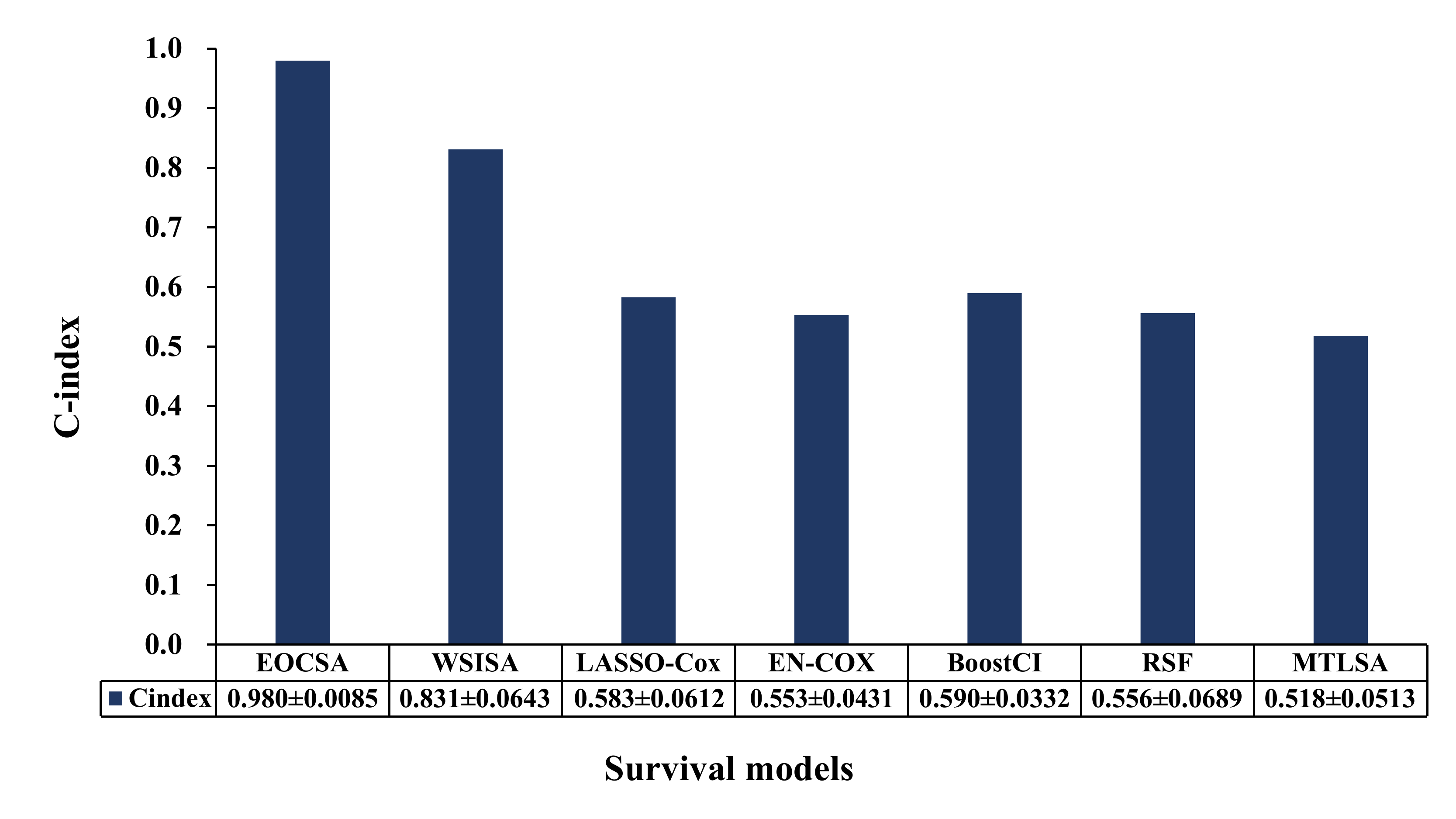}
	\caption{The results of EOCSA framework and other frameworks. It shows that the C-index value of our proposed EOCSA is higher than other frameworks.}
	\label{ComparisonWithOthers}
\end{figure}

The results of different survival models in the EOCSA and other survival frameworks are shown in Fig.~\ref{ComparisonWithOthers}. It shows that EOCSA framework achieves higher accuracy than other survival frameworks. The C-index value of EOCSA is $ 0.149 $, $ 0.397 $, $ 0.427$, $ 0.390 $, $ 0.424 $, and $ 0.462 $ higher than WSISA, F$_{hand}$+LASSO-COX, F$_{hand}$+EN-COX, F$_{hand}$+BoostCI, F$_{hand}$+RSF, and F$_{hand}$+MTLSA respectively. These differences illustrate the power of deep learning in feature extraction and the strong survival prediction ability of EOCSA.

\subsection{Generalization Evaluation}
\begin{figure}[t]
	\centering
	\subfigure[One year]{
		\includegraphics[width=5.5cm]{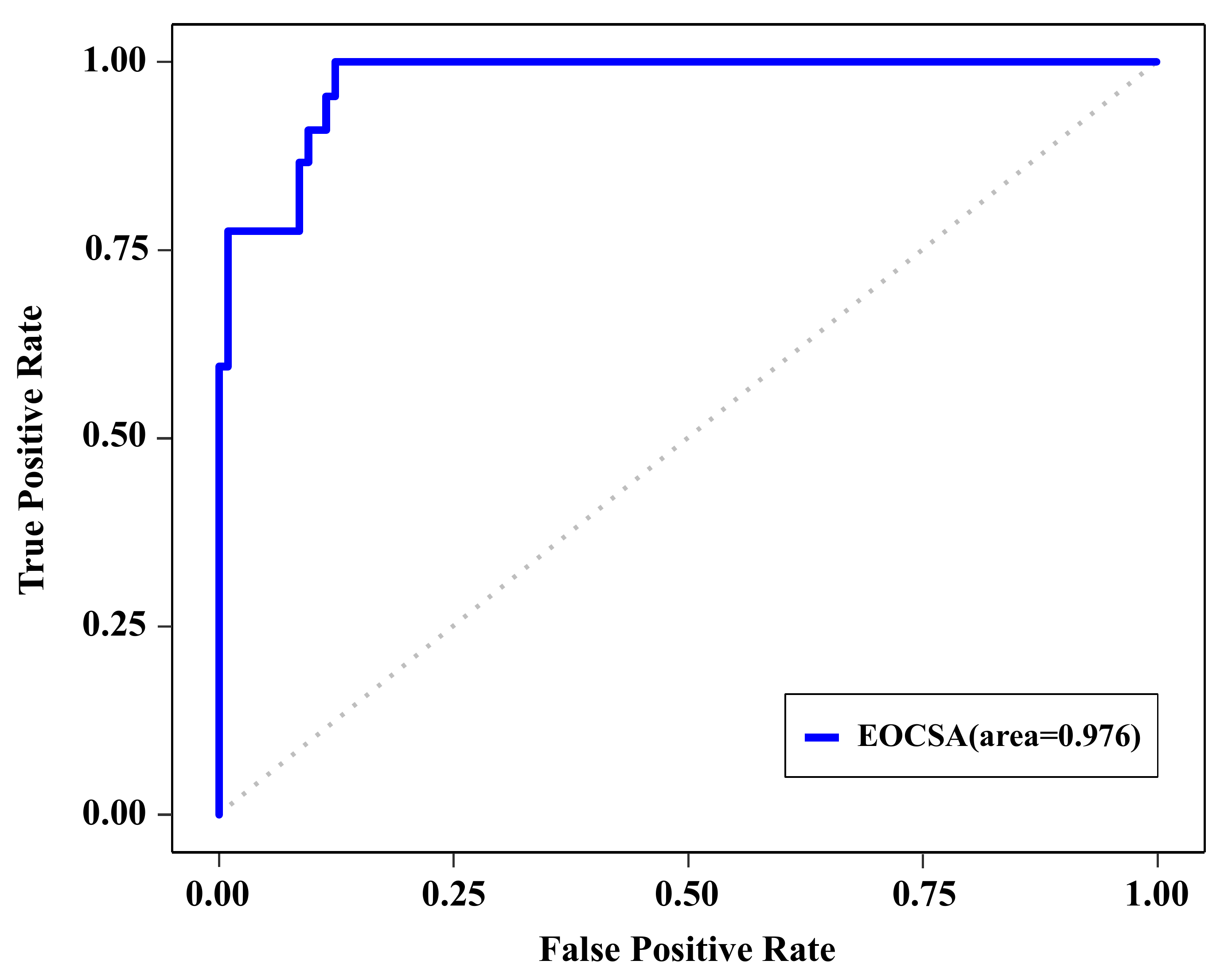}
	}
	\subfigure[Three years]{
		\includegraphics[width=5.5cm]{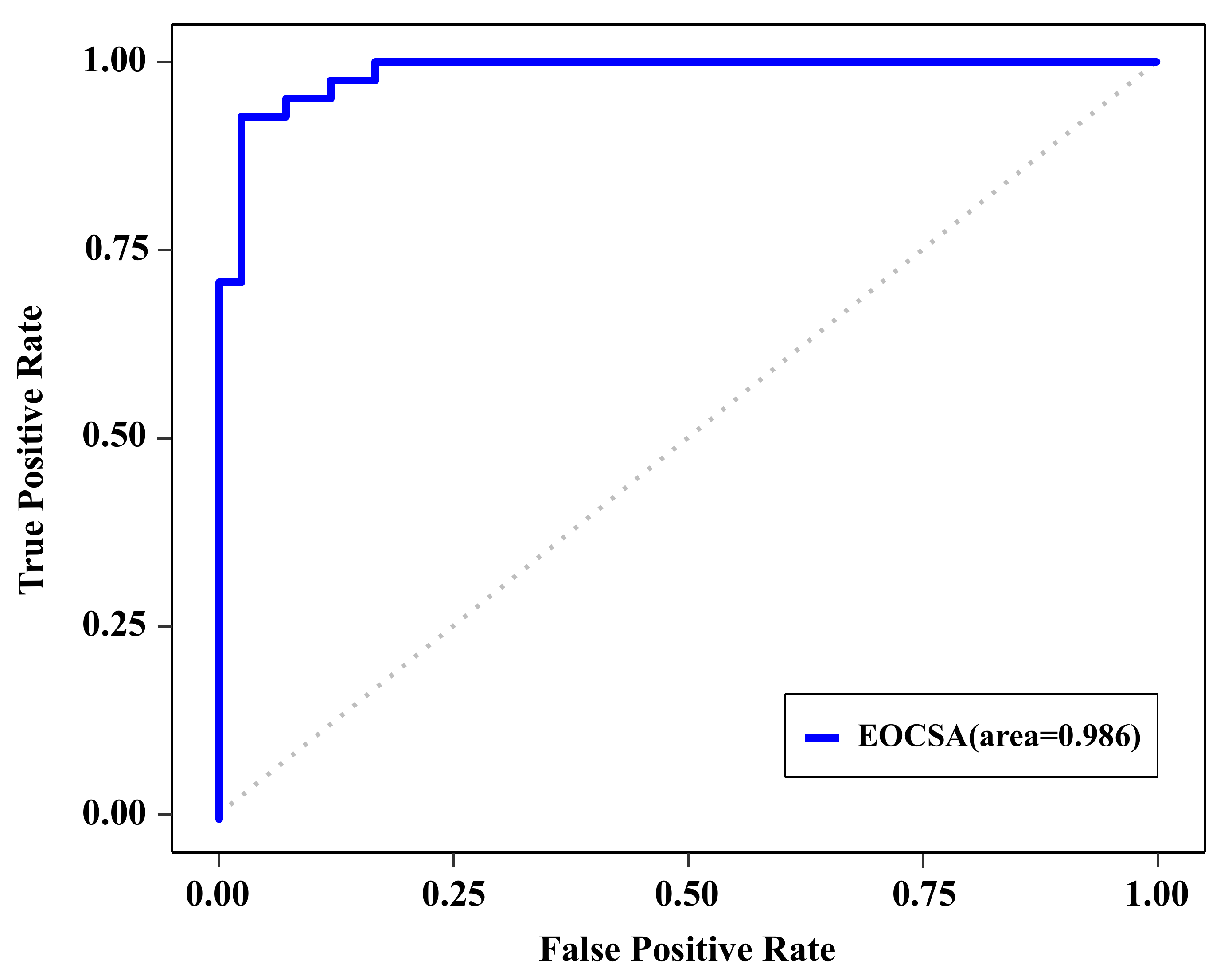}
	}
	\subfigure[Five years]{
		\includegraphics[width=5.5cm]{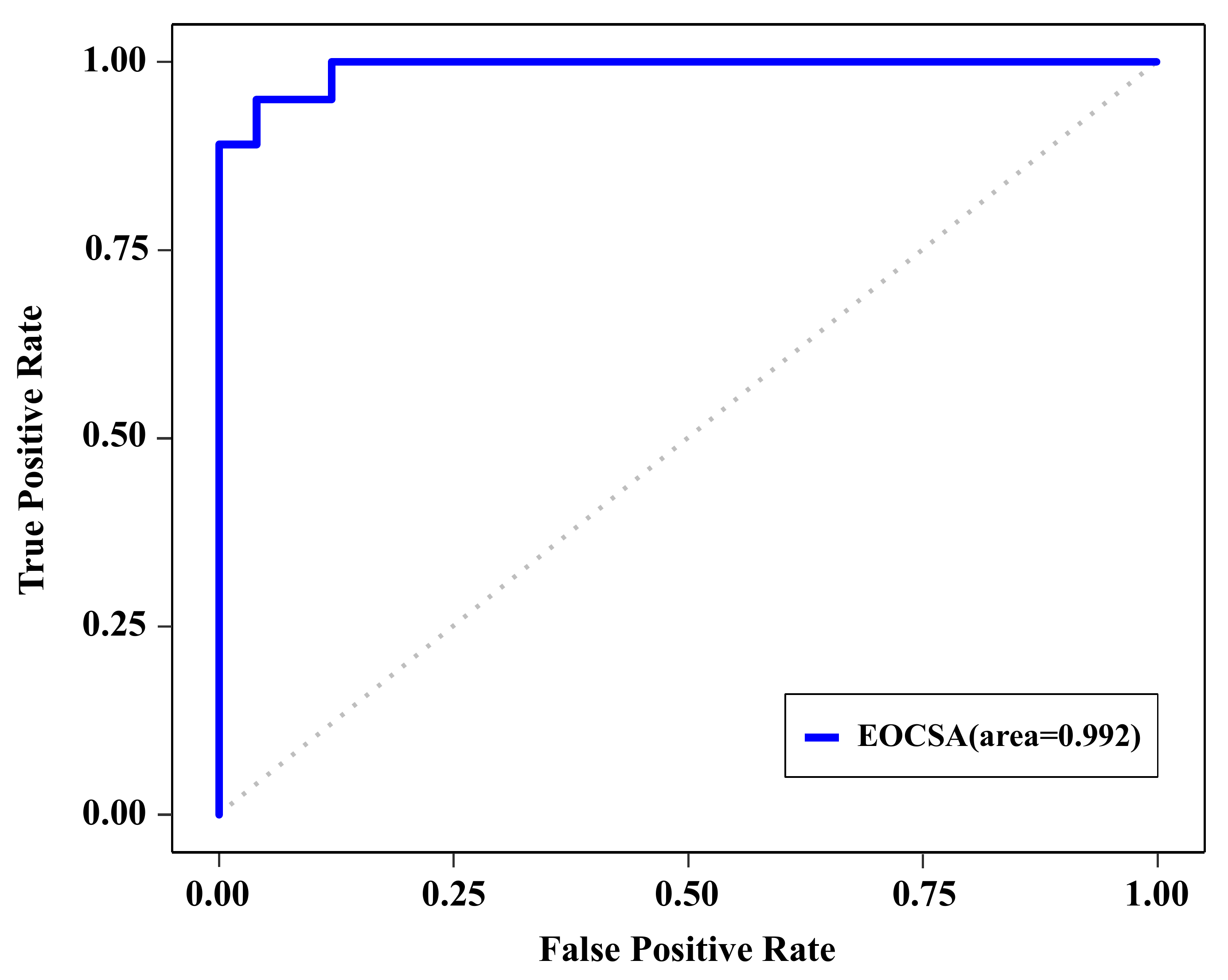}
	}
	\subfigure[High-low risk prediction]{
		\includegraphics[width=5.8cm]{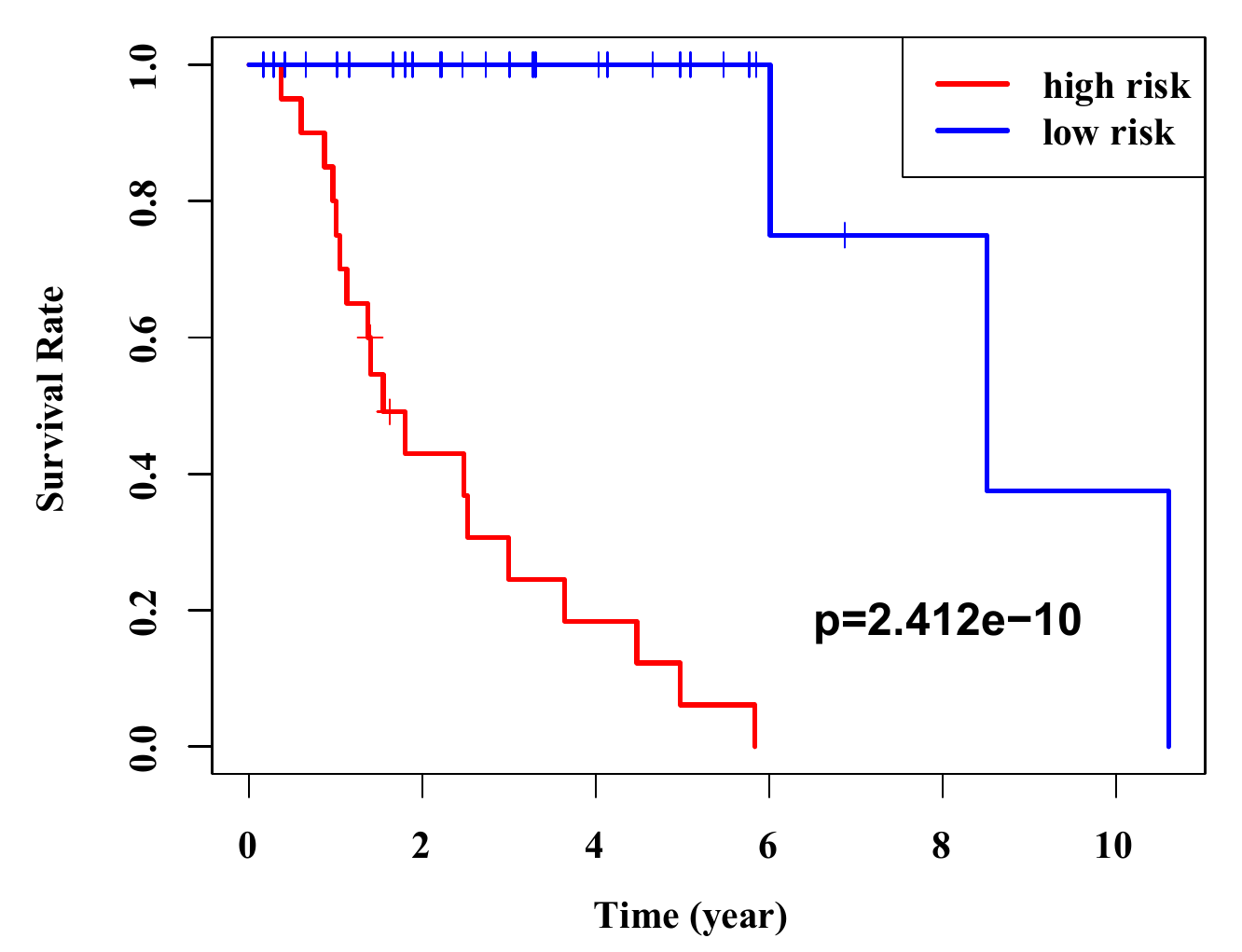}
	}
	
	\caption{The comparison of EOCSA and WSISA survival ROCs when setting with different prediction time.}
	\label{evaluationWithLUSC}

\end{figure}

Furthermore, we also evaluation our proposed EOCSA in TCGA-LUSC dataset for proving the generalization. There are 498 patients and 1265 WSIs being selected for experiments. Our proposed EOCSA obtains $0.966\pm0.0096$ C-index value. The time-dependent survival ROCs and high-low risk prediction ROC are shown in Fig.~\ref{evaluationWithLUSC}. The AUC values are 0.976, 0.986 and 0.992 with prediction time being set to 1, 3 and 5 years, which are shown in Fig.~\ref{evaluationWithLUSC}(a), Fig.~\ref{evaluationWithLUSC}(b) and Fig.~\ref{evaluationWithLUSC}(c) respectively. Fig.~\ref{evaluationWithLUSC}(d) shows the survival curve for high-low risk prediction and the proposed EOCSA obtains p-value with $2.412e^{-10}$. The above-mentioned results on TCGA-LUSC have demonstrated that the proposed EOCSA has good generalization performance.

\subsection{Implementation Details}
The proposed EOCSA framework was implemented in the Python environment (Python foundation version 3.5.2), R environment (R foundation version 3.5.3), and Windows 10 operating system (Microsoft, USA). The proposed DCAS model was built using the ``tensorflow-gpu'' package. LASSO-Cox was built using the ``coxph'' function from the ``glmnet'' package~\citep{glmnet}, and EN-Cox was built with the ``cocktail'' function in the ``fastcox'' package~\citep{Cocktail}. RSF was implemented with the ``randomForestSRC'' package (https://github.com/kogalur/randomForestSRC). The supplementary materials of~\citep{mayr2014boosting} shared the code of BoostCI.  The code of MTLSA could be downloaded from ``https://github.com/yanlirock/MTLSA''. The survival ROC curve was plotted using the ``timeROC'' package and the survival curves of high risk and low risk were generated by the ``survival'' package. The hand-crafted features were extracted by CellProfiler (version 3.1.9)~\citep{carpenter2006cellprofiler}.

\section{Conclusion}
In this study, we proposed a EOCSA framework which focused on the survival analysis of EOC by processing WSIs using deep leaning technologies. We have developed a prediction model named DCAS and the experimental results show the outstanding feature extraction ability and strong predictive power. The DCAS adopted a ``cluster-selection strategy'' to efficiently remove redundant information. Different from the other deep survival prediction models, we embedded spatial and channel attention modules to capture tumor information. Our weight calculation method also enables the generation of more discriminative patient-level features.

We have compared with the EOCSA and the DCAS with existing methods such as WSISA and deepConvSurv, and achieved state-of-the-art results. It is worth noting that the EOCSA framework obtained a C-index value of $0.980$ which was $0.123$ higher than the WSISA, which is a general approach for predicting survival based on WSIs. However, there are still some limitations in EOCSA framework. Firstly, cluster selection is effective to choose the discriminative patches, but it is quite costly in calculation. The efficiency of the EOCSA framework will decrease if extremely huge number of WSIs are feeded in to the model; Secondly, not all chosen patches are sufficiently discriminative. It will affect the survival analysis results if too many non-discriminative patches are chosen. In the future, we aim to propose more efficient and effective patch selection approach. And our model is expected to be employed in other types of cancers for prognosis analysis.

\section*{Acknowledgment}

This study was supported by the National Natural Science Foundation of China (Grant Nos. 62072329 and 62071278).\vspace*{-12pt}

\section*{References}

\bibliography{reference_6}

\end{document}